\documentclass[12pt]{article}

\usepackage{latexsym,amsmath,amssymb,theorem,epsfig}

\DeclareFontFamily{OT1}{rsfs10}{}
\DeclareFontShape{OT1}{rsfs10}{m}{n}{ <-> rsfs10 }{}
\DeclareMathAlphabet{\mathscript}{OT1}{rsfs10}{m}{n}

\numberwithin{equation}{section}

\newcommand{\ns}{\normalsize}

\def\cO{{\mathcal O}}

\def\cV{{\mathcal V}}

\def\gsim{ \lower .75ex \hbox{$\sim$} \llap{\raise .27ex \hbox{$>$}} }
\def\lsim{ \lower .75ex \hbox{$\sim$} \llap{\raise .27ex \hbox{$<$}} }
\def\be{\begin{equation}}
\def\ee{\end{equation}}
\def\bea{\begin{eqnarray}}
\def\eea{\end{eqnarray}}

\theoremstyle{plain}

{\theorembodyfont{\rmfamily} }

\begin{document}


\begin{titlepage}

\vspace{-5cm}

\title{
  \hfill{\ns }  \\[1em]
   {\LARGE New Ekpyrotic Cosmology}
\\[1em] }
\author{
   Evgeny I. Buchbinder$^{1}$, Justin Khoury$^{1}$, Burt A. Ovrut$^{2}$
     \\[0.5em]
   {\ns ${}^1$ Perimeter Institute for Theoretical Physics} \\[-0.4cm]
{\ns Waterloo, Ontario, N2L 2Y5, Canada}\\[0.3cm]
{\ns ${}^2$ Department of Physics} \\[-0.4cm]
{\ns The University of Pennsylvania}\\[-0.4cm]
{\ns  Philadelphia, PA 19104--6395, USA}\\[0.3cm]}

\date{}

\maketitle

\begin{abstract}
In this paper, we present a new scenario of the early Universe that contains
a pre big bang Ekpyrotic phase. By combining this with a ghost condensate, 
the theory explicitly violates the null energy condition without developing any ghost-like instabilities.
Thus the contracting universe goes through a non-singular bounce and evolves smoothly into the expanding
post big bang phase. The curvature perturbation acquires a scale-invariant spectrum well before the bounce in this scenario.
It is sourced by the scale-invariant entropy perturbation engendered by two ekpyrotic scalar fields, a mechanism recently
proposed by Lehners {\it et al.} Since the background geometry is non-singular at all times, the curvature perturbation remains nearly constant
on super horizon scales. It emerges from the bounce unscathed and imprints a scale-invariant spectrum of density fluctuations in the
matter-radiation fluid at the onset of the hot big bang phase.  
The ekpyrotic potential can be chosen so that the spectrum has a ``red'' tilt, in accordance 
with the recent data from WMAP. 
As in the original Ekpyrotic scenario, the model predicts a negligible 
gravity wave signal on all observable scales.
As such ``New Ekpyrotic Cosmology" provides a 
consistent and distinguishable alternative to
inflation to account for the origin of the seeds of large scale structure.
\end{abstract}

\thispagestyle{empty}

\end{titlepage}

\section{Introduction}

Over the past decade observations of the microwave background temperature anisotropy have revealed that the
large scale structure in our universe originates from primordial perturbations that are nearly scale-invariant, adiabatic and
gaussian. Since these coincide with the predictions of the simplest inflationary models, this is widely regarded as evidence 
for inflation. However this does not constitutes a proof, and it is prudent to keep in mind that the seeds for structure formation could originate
from a different mechanism. Ultimately our faith in inflation must rely on the absence of a compelling alternative paradigm.

In this paper, we present a fully consistent and complete scenario of early universe cosmology which produces a nearly scale-invariant spectrum
of density fluctuations without invoking a period of accelerated expansion. The model is strongly inspired by and borrows key ingredients from ekpyrotic~\cite{ek1,seiberg,ekpert} and cyclic~\cite{cyc} cosmology, in particular the idea that density perturbations are generated during a slow contracting phase prior to the big bang.
A key difference, however, is that the cosmological evolution is now completely non-singular and avoids any big crunch singularity. Moreover, perturbations remain in the linear regime
throughout. Thus, entirely at the level of a 4d effective theory, we are able to produce a scale-invariant spectrum of density fluctuations, track its evolution through
the reversal from contraction to expansion, and its transmission to the matter-radiation fluid at the onset of the hot expanding phase. The idea of a contracting phase prior to the big bang originated in pre-big bang cosmology. Density perturbations and the evolution from the contracting to the expanding phase have been discussed in this context. See~\cite{veneziano} for a review.

The key ingredient in ekpryosis is a scalar field rolling down a negative, nearly exponential potential. If the coefficient in the exponent is sufficiently large,
then, as the scalar field rolls down the potential, its fluctuations acquire a nearly scale-invariant spectrum spanning 60 e-foldings in comoving wavenumber~\cite{ek1,compare}.
Since the potential energy is negative, this occurs in a slowly contracting 
Friedmann-Lemaitre-Robertson-Walker (FLRW) geometry and, hence, with a rapidly
decreasing Hubble radius. In some sense this ekpyrotic phase is  ``dual'' to 
inflationary cosmology~\cite{compare,dual}, which, in contrast, is a period of 
exponential expansion with nearly constant Hubble radius. 
The difference in dynamics, however, leads to a distinguishing observational 
prediction: the inflationary gravitational wave spectrum is nearly scale-invariant, 
whereas that of ekpyrosis is not~\cite{ek1,gwaves}. In ekpyrotic theories, the gravitational wave 
spectrum is strongly blue and, hence, the amplitude is exponentially suppressed on all 
observable scales~\cite{gwaves}.

An important obstruction facing ekpyrotic theory is how to ``bounce'' from the contracting
phase to an expanding phase, which requires a violation of the null energy condition (NEC).
This is no small feat, however, since non-singular theories that violate the NEC generally suffer from violent instabilities, such as
ghosts or tachyons of arbitrarily large mass~\cite{nic}. In the Big Bang/Big Crunch ekpyrotic scenario of~\cite{seiberg, ekpert}, as well as in the cyclic model~\cite{cyc}, one therefore allows the FLRW space-time to crunch, invoking 
stringy effects to effectively violate the NEC and generate a smooth bounce. This is motivated by the relative
mildness of the singularity~\cite{seiberg}. (Understood, for example, as a collision of end-of-the-world branes in heterotic $M$-theory~\cite{het1, het1A}, only the fifth dimension shrinks to zero size while the 3 large 
dimensions remain finite.) Despite considerable effort in string theory~\cite{bunch} and compelling physical arguments~\cite{perry}, there is still no proof that a bounce is possible. 

Recently, however, NEC-violating solutions~\cite{Ghost3} have been derived in the context of ghost condensation~\cite{Ghost1,Ghost2}. Since these models involve higher-derivative kinetic terms, they evade the assumptions of~\cite{nic} and therefore yield ghost-free solutions. 

In this paper, we show explicitly how the NEC-violating ghost condensate can be merged consistently with the preceding ekpyrotic phase to generate a non-singular bouncing cosmology. This merger involves many subtleties. For instance, since the energy density of an NEC-violating fluid grows in an expanding universe, whereas everything else redshifts,  it quickly comes to dominate the universe. But in a contracting universe precisely the opposite happens: the NEC-violating component goes to zero while any amount of radiation or scalar kinetic energy blueshifts. Thus achieving a bounce hinges on an efficient transfer of energy from the ekpyrotic scalar, which dominates during the ekpyrotic phase, to the ghost condensate, which achieves the bounce. To realize this we therefore propose that they are, in fact, one and the same field. The higher-derivative kinetic function is chosen to be nearly canonical during the ekpyrotic phase, while higher derivatives become relevant only as we approach the bounce. This translates into consistency relations between the kinetic function and the scalar potential, which we derive explicitly. We argue that these conditions are realized for a wide class of kinetic functions and scalar potentials.

Armed with a non-singular cosmological evolution, we can address the propagation of density perturbations through the bounce, an issue that has stirred a lot of controversy over the past few years~\cite{lythandfriends}. Although the fluctuations in the scalar field and the gauge-invariant Newtonian potential are scale-invariant, 
it turns out that their growing mode precisely cancels out of $\zeta$ --- the curvature perturbation on uniform-density hypersurfaces~\cite{bardeen}. The latter is a useful variable to track since it is conserved on super-horizon scales, in the absence of entropy perturbations. Thus, as long as 4d general relativity is valid, $\zeta$ remains constant independent of the physics of the bounce. In the context of the Big Bang/Big Crunch ekpyrotic or cyclic models, however, 4d gravity must break down near the singularity and therefore the issue of what happens to $\zeta$ remains unresolved. Some maintain that the likely matching condition is for $\zeta$ to be continuous, most convincingly~\cite{paolonic}, while others argue that higher-dimensional effects near the singularity could lead to mode-mixing and endow $\zeta$ with a scale-invariant contribution~\cite{tolley,mcfadden,other5d}. Nevertheless, one must also deal with the fact that this mode diverges logarithmically near the singularity, resulting in a breakdown of perturbation theory, although matching prescriptions based on analytic continuation have been proposed~\cite{tolley,tolleybefore}.

In our case, the evolution is completely non-singular and, hence, 
the outcome of the perturbations is unambiguous: $\zeta$ is conserved. 
To generate a scale-invariant spectrum we rely instead on a recently proposed 
mechanism using entropy perturbations~\cite{private,talks}. For earlier and closely 
related work, see~\cite{notari,finelli1,finelli2,finelli3}. If we have two scalar fields, 
each rolling down a steep, negative, and nearly exponential potential, then each of them 
acquires a scale-invariant spectrum of fluctuations, as described earlier. 
Moreover, the entropy perturbation --- corresponding to the difference in the scalar 
fluctuations --- is also scale-invariant~\cite{private,finelli1}. By converting this 
entropy perturbation into the adiabatic mode ({\it i.e.}, $\zeta$), the curvature 
perturbation thus becomes scale-invariant long before the bounce.
In the inflationary context, the idea of using a spectator field to generate a scale-invariant spectrum 
of entropy perturbations, later to be imprinted on the adiabatic mode, was also considered in the curvaton~\cite{C1} and
modulon scenarios~\cite{C2}.

In this paper, we propose a concrete realization of the entropy generation mechanism. We consider two ekpyrotic scalar fields, each with its own negative potential and its
own higher-derivative kinetic function. As the fields roll down their respective potentials, the entropy perturbation acquires a scale-invariant spectrum.
To convert this spectrum into the adiabatic mode, we assume that the ekpyrotic potentials are such that one field exits the ekpyrotic phase and enters the ghost condensation regime before the other. More precisely, the potential in the ekpyrotic phase is steep and negative, as mentioned earlier, while in the ghost condensate phase it must be positive. Thus this transition is marked by a sharp rise in the potential. If the potentials are such that one field hits the transition before the other, this will result in a sharp turn in the trajectory in field space, and in the process $\zeta$ will acquire a scale-invariant piece from the entropy perturbation. We calculate the resulting $\zeta$ explicitly and find an expression that closely resembles its inflationary counterpart. This confirms earlier claims that the required level of tuning on the ekpyrotic potential is comparable to that in inflation.

For {\it exact} exponential potentials, we find that the resulting spectral tilt is slightly ``blue", which at first sight is disturbing in light of
the recent WMAP evidence for a small ``red" tilt. However, we show that deviations from the pure exponential form can lead to a small red tilt.
See also~\cite{private} for an independent derivation. This is shown explicitly in the case that the potentials for the two scalar fields are identical, albeit not exactly exponentials. This simplification is made only to facilitate the analysis, and we do not believe that the resulting red tilt hinges on it. In this limit we derive an exact evolution equation for the entropy perturbation, cast entirely
in terms of the background equation of state, which closely resembles analogous results in inflation~\cite{wang} and old (single-field) ekpyrotic theory~\cite{gratton}. The resulting spectral tilt
acquires a dependence on the degree of departure from pure exponential form, and, in particular, can be red. 

Coming back to the background evolution, one of the potential dangers with contracting universes is that the FLRW background is unstable to the
onset of chaotic mixmaster behavior~\cite{BKL,chaosothers}. In~\cite{joel} it was shown that mixmaster behavior is suppressed if the dominant
energy component has equation of state $w\gg 1$, as is the case in the ekpyrotic contracting phase. 
In singular ekpyrotic theories, as the universe approaches the singularity the equation
of state must eventually revert to $w=1$, and mixmaster behavior can potentially resurface. A necessary condition to avoid a chaotic bounce is that the anisotropy and curvature
components be exponentially suppressed at the onset of the $w=1$ phase so that they remain subdominant all the way to the string or Planck scale, where 4d gravity breaks down anyway. 

Mixmaster issues also apply to our new scenario but are trivially satisfied since the bounce is non-singular. The $w\gg 1$ ekpyrotic phase exponentially suppresses anisotropy and curvature components. The latter grow again during the NEC-violating phase, but only by an insignificant amount since the bounce occurs within one e-fold of contraction.

To summarize, our scenario is a complete and unambiguous template for early universe cosmology.
It provides an alternative explanation for the origin of the seeds of structure formation
in which perturbations are generated long before the bounce, all within a singularity-free and finite FLRW background.
The bounce is entirely described at the level of a consistent effective theory, which is free of ghost-like instabilities or other pathologies.
Mixmaster behavior is trivially avoided. Perturbations remain in the linear regime, and their evolution can be tracked through the bounce. Since 4d gravity remains valid throughout, the curvature perturbation $\zeta$ goes through unscathed. And since $\zeta$ acquires a scale-invariant spectrum long before the bounce, thanks to the entropy perturbation conversion mechanism, it is unequivocally
scale-invariant after the transition. The universe therefore emerges in a hot big bang phase, endowed with a super-horizon spectrum of scale-invariant fluctuations.
While the approach presented here is closely related to and borrows key elements of the original ekpyrotic scenario~\cite{ek1}, our new scenario resolves all of its short-comings, such as generating a
bounce. It also resolves the important issues of singularity-avoidance
and the fate of the perturbations that occurs in~\cite{seiberg, ekpert, cyc}. For these reasons, we call it  ``New Ekpyrotic Cosmology''.

As it stands the new ekpyrotic scenario is precisely that, a scenario. 
In principle, it might be implemented in various ways in different fundamental theories of particle 
physics. That being said, this scenario maintains the motivations of the 
original ekpyrotic model~\cite{ek1}, namely, as the cosmology associated 
with the singularity free collision of a bulk five-brane~\cite{het2, het2A}  in heterotic 
M-theory with the observable boundary wall. Indeed, we have computed the 
potential in such a theory and find that it can satisfy many of the 
constraints required in the new ekpyrotic model. This potential will 
be presented elsewhere~\cite{future}.  A further motivation is that a realistic matter spectrum
can appear on the observable wall of such theories, see, {\it e.g.},~\cite{mssm}. Hence, our scenario can potentially occur in a realistic context.
More generally, we emphasize that, with the exception of the so-called no-scale theories,  a
supersymmetric minimum in the scalar space of any 4d $N=1$ supergravity 
theory, including effective low energy superstring theories, has a negative 
cosmological constant. This can remain true in the presence of flux and
non-perturbative effects, even for a vacuum that breaks supersymmetry~\cite{het3}. Therefore, potentials consistent with  
ekpyrotic cosmology, {\it i.e.}, where scalars roll down a steep negative potential energy, appear naturally in this context.

A more pressing question is whether a ghost
condensate can occur in string theory. It has been argued that this is impossible given the known
analytic properties of the string S-matrix~\cite{IRnima}. See ~\cite{Mukohyama}, however, for a recent attempt to overcome these difficulties.
Without question higher derivative interactions do occur in string theories, but whether they are of the requisite form is an 
open issue. Ultimately a non-singular bounce might arise from very different physics, unrelated to string theory. In fact, the NEC could conceivably be smoothly violated by a different mechanism than ghost condensation. Be that as it may, many of our results would continue to apply to any such mechanism.
The rationale for focusing here on the ghost condensate is to provide an explicit and complete realization of our new scenario.

In Sec.~\ref{1field} we review the essential concepts of the ekpyrotic phase, as a theory of a scalar field rolling down a negative potential. We show how the fluctuations
in this field acquire a scale-invariant spectrum, which unfortunately projects out of the curvature perturbation. In Sec.~\ref{2fields} we extend the discussion to two scalar fields and review the mechanism proposed by~\cite{private,talks} to generate a scale-invariant spectrum for the entropy perturbation. In Sec.~\ref{red} we derive an explicit expression for the spectral tilt and argue that it can be slightly red, in agreement with recent microwave background data. We show in Sec.~\ref{imprint} how this gets imprinted into the curvature perturbation by requiring that the two fields exit the ekpyrotic phase at different times. Next we turn to the bounce, first, in Sec.~\ref{remarks}, discussing
general requirements on its physics followed by a brief review of ghost condensation in Sec.~\ref{ghostcond}. In Sec.~\ref{merge} we show in detail that the ekpyrotic phase and ghost condensate can be merged successfully to generate a non-singular cosmological scenario, and derive consistency requirements on the kinetic function and scalar potential. Section~\ref{reheat} presents a short discussion of the physics of reheating in this scenario, while Sec.~\ref{conclude} provides some concluding remarks.

\section{Review of Single-Field Ekpyrosis}
\label{1field}

At the level of a 4d effective description, the basic ingredients of the simplest ekpyrotic scenario are essentially the same as in inflation, namely a scalar field $\phi$ rolling down some self-interaction potential $V(\phi)$. A key difference, however, is that while inflation requires a flat and positive potential, its ekpyrotic counterpart is {\it steep} and {\it negative}. This has a dramatic impact on the cosmological evolution. Instead of accelerated expansion, an ekpyrotic theory has slow contraction. Instead of an exponentially growing scale factor and nearly constant Hubble radius, corresponding to approximate de Sitter geometry, we now have a nearly constant scale factor and rapidly shrinking Hubble radius, corresponding to approximately flat space. 

\subsection{Ekpyrotic Potential}

\begin{figure}[ht]
\centering
\includegraphics[width=100mm]{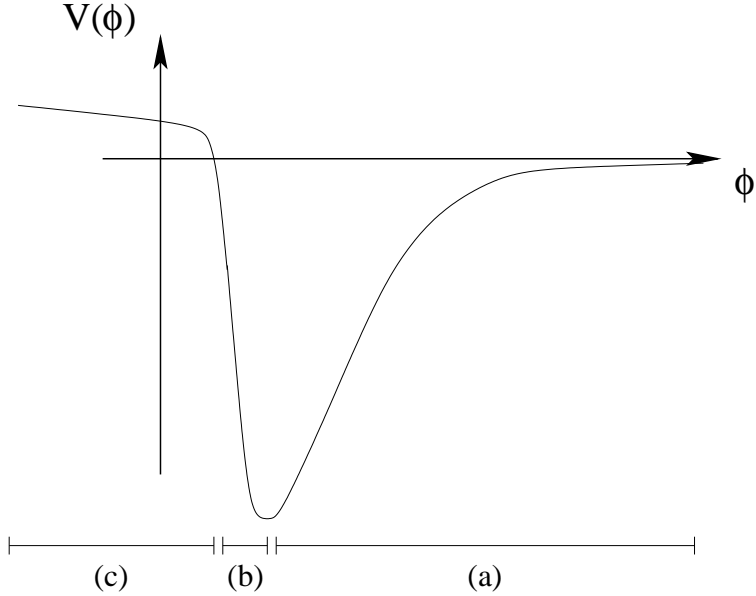}
\caption{Generic shape of the ekpyrotic scalar potential. }
\label{pot}
\end{figure}

A generic ekpyrotic potential, shown in Fig.~\ref{pot}, consists qualitatively of three parts. In the region denoted by~(a), the potential must be steep and negative. As the field rolls down this part of the potential, there is a scaling solution, which is an attractor, corresponding to very slow contraction. 

It is also during this phase that large-scale density fluctuations are generated as modes 
exit the horizon. As we will review later, in order for the spectrum to be nearly scale-invariant,  
the potential must satisfy the ``fast-roll'' conditions~\cite{design}:
\be
\epsilon\ll 1\;;\qquad |\eta|\ll 1\,,
\ee
where
\be
\epsilon \equiv  M_{\rm Pl}^{-2}\left(\frac{V}{V_{,\,\phi}}\right)^2\,; \qquad \eta \equiv   1-\frac{V_{,\,\phi\phi}V}{V_{,\,\phi}^2} \,,
\label{fastroll}
\ee
%
are fast-roll parameters, in analogy with the standard slow-roll parameters in
inflation\footnote{Recall that the slow-roll parameters in inflation are 
$\epsilon_{{\rm inf}}=\frac{M_{{\rm Pl}}}{2} \left(\frac{V_{, \phi}}{V}\right)^2$ and
$\eta_{{\rm inf}}=M_{{\rm Pl}}^2\left(\frac{V_{, \phi \phi}}{V}\right)$.}.
Here $V_{, \phi}\equiv dV/d\phi$, and $M_{{\rm Pl}}$ is the ``reduced'' Planck mass: $M_{{\rm Pl}}=2.4 \times 10^{18}$ GeV.
These conditions respectively require the potential to be steep and nearly exponential, and thus region~(a) of the potential can be approximated by
\be
V(\phi)\approx -V_0\exp\left(-\sqrt{\frac{2}{p}}\frac{\phi}{M_{\rm Pl}}\right)\,,
\label{expneg}
\ee
with $p\ll 1$. 

Much as  inflation ends when the flatness condition breaks down, here as well the scaling behavior terminates once equations~(\ref{fastroll}) are no longer satisfied. In order to avoid being left with a large negative vacuum energy at the end of the ekpyrotic phase, let us assume that the potential has a minimum and rises back up towards positive values, as shown in region~(b) of Fig. 1. 

There is considerable freedom in specifying the shape of the potential further to the left of the minimum, 
that is, in region (c). Since it is in this region that the universe reverses from contraction to expansion, the detailed shape of the potential is dependent on the explicit mechanism producing the bounce. In this paper, we will use ghost condensation to generate a non-singular bounce where all instabilities are under control. As we will see in Sec.~\ref{ghostcond}, this requires a flat and positive potential in region (c), as sketched in Fig.1.

\subsection{Scaling Solution}

The ekpyrotic phase occurs as the field rolls down the steep, negative exponential part of the 
potential; region (a) in Fig.~\ref{pot}. In this paper, we take the 
background geometry to be a homogeneous, isotropic Friedmann-Lemaitre-Robertson-Walker (FLRW) space-time 
which is spatially flat. That is,
\begin{equation}
ds^{2}=-dt^{2}+a(t)^{2}{d\vec{x}}^{2}.
\label{burt1}
\end{equation}
Furthermore, in this section, $\phi$ is chosen to have canonical kinetic energy and no higher 
derivative interactions. The equations of motion are then given as usual by the Friedmann equation,
\be
3H^2M_{\rm Pl}^2 = \frac{1}{2}\dot{\phi}^2 + V(\phi)\,,
\ee
as well as the evolution equation for the scalar field:
\be
\ddot{\phi} + 3H\dot{\phi} = -V_{,\,\phi}\,.
\label{phieom}
\ee

For the potential~(\ref{expneg}), it is easily seen that these equations allow for an exact scaling solution~\cite{ekpert,gratton,paolonic}
\bea
\nonumber
& & a(t)\sim (-t)^p\,;\qquad H = \frac{p}{t}\,; \\
& & \phi(t) = \sqrt{2p}M_{\rm Pl}\log\left(-\sqrt{\frac{V_0}{M_{\rm Pl}^2p(1-3p)}}\; t\right)\,,
\label{scaling1}
\eea
where $t$ is negative and increases towards zero. Hence the solution describes a slowly contracting universe. Since $p\ll 1$, we see that the kinetic and potential energy of the scalar are both large in magnitude, but nearly cancel each other to yield a small total energy density. In other words, $\dot{\phi}^2/2 \gg H^2M_{\rm Pl}^2$, and $|V|\gg H^2M_{\rm Pl}^2$.
 
This scaling solution, moreover, has the desirable property that it is an attractor. Indeed, the scalar field has the equation of state of a very stiff fluid:
\be
w\equiv \frac{\cal{P}}{\rho} = \frac{2}{3p}-1\gg 1\,.
\ee
This means that its energy density behaves as $\rho_\phi\sim a^{-2/p}$. But since $p\ll 1$, this energy component {\it blueshifts} much more rapidly than any other relevant contribution to the Friedmann equation. Any curvature ($\sim a^{-2}$), matter ($\sim a^{-3}$), radiation ($\sim a^{-4}$), or anisotropy ($\sim a^{-6}$) component, for instance, quickly becomes subdominant to the scalar field energy density. Incidentally, this is precisely analogous to inflation where the scalar potential gives a nearly constant source in the Friedmann equation. Such a term therefore redshifts much more slowly than any of the aforementioned relevant contributions, which is why accelerated expansion is an attractor.

\subsection{Density Perturbations --- Newtonian Potential Analysis} \label{newton}

By far the most important feature of a negative exponential potential, however, is that it generates a scale-invariant spectrum of fluctuations in the scalar field, even when the gravitational interactions are turned off. 
Indeed, in the absence of gravity,~(\ref{phieom}) can be integrated trivially:
\be
\dot{\phi} = -\sqrt{-2V}\,,
\ee
where we have set the total energy to zero. For the pure exponential potential~(\ref{expneg}),
the solution is
\be
-t = \int_{-\infty}^\phi d\tilde{\phi} \frac{e^{\tilde{\phi}/\sqrt{2p}M_{\rm Pl}}}{\sqrt{2V_0}}  
   = M_{\rm Pl}\sqrt{p}\frac{e^{\phi/\sqrt{2p}M_{\rm Pl}}}{\sqrt{V_0}} = \sqrt{\frac{2}{-V_{,\,\phi\phi}}}\,.
\label{tsoln}
\ee

Now fluctuations in the scalar field satisfy the perturbed equation
\be
\ddot{\delta\phi}_k + \left(k^2 + V_{,\,\phi\phi}\right)\delta\phi_k = 0\,,
\ee
which, using~(\ref{tsoln}), can be written as
\be
\ddot{\delta\phi}_k + \left(k^2 - \frac{2}{t^2}\right)\delta\phi_k = 0\,.
\label{pert}
\ee
This describes a harmonic oscillator with a time-dependent mass. Assuming standard Bunch-Davies 
boundary conditions~\cite{BD}, 
$\delta\phi_k\rightarrow e^{ikt}/\sqrt{2k}$ as $k(-t)\rightarrow \infty$, the general solution for the mode functions is given by a Hankel function of the first kind: $\delta\phi_k\sim \sqrt{-t}H_{3/2}^{(1)}(-kt)$. In the long-wavelength limit, $k(-t)\rightarrow 0$, this gives
\be
\delta\phi_k^2\sim k^{-3}\,,
\ee
corresponding to a scale-invariant power spectrum. Scale invariance can be traced back to the factor of 2 coefficient in the time-dependent term in~(\ref{pert}). Remarkably, from~(\ref{tsoln}) we see that this holds for any $p$. When we include gravitational interactions, however, we will find that scale invariance requires $p\ll 1$.

Turning on gravity, the gauge-invariant variable that faithfully reproduces the above scale-invariant spectrum is the Newtonian potential $\Phi$, the scalar metric fluctuation in Newtonian gauge. It is convenient to do the analysis in terms of a related gauge-invariant variable, $u\equiv a\Phi/\phi'$, where primes denote differentiation with respect to conformal time $\tau$. Its Fourier modes satisfy~\cite{ekpert}
\be
u_k'' + \left(k^2 -\frac{p}{(1-p)^2\tau^2}\right)u_k=0\,
\ee
whose solution is once again given by a Hankel function
\be
u_k = \frac{\sqrt{p}}{(2k)^{3/2}M_{\rm Pl}}\sqrt{\frac{\pi}{2}}\sqrt{-k\tau}H_\nu^{(1)}(-k\tau)\,,
\label{usoln}
\ee
where $\nu \equiv (1+p)/2(1-p)$, and we have assumed Bunch-Davies initial conditions: $u_k\rightarrow e^{-ik\tau}/(2k)^{3/2}$. The power spectrum on large scales is then
\be
k^3u_k^2\sim k^{-2p/(1-p)}\,,
\ee
which is indeed scale-invariant if $p\ll 1$. The spectral index in this limit is $n_s-1\approx -2p$, corresponding to a slight red tilt. For more general potentials, the expression for $n_s$ in terms of the fast roll parameters~(\ref{fastroll}) closely resembles its slow-roll counterpart in inflation:
\be
n_s - 1 = -4(\epsilon + \eta)\,.
\label{single}
\ee
In particular, potentials with non-trivial $\eta$, characterizing deviations from pure exponential, can lead to a blue tilt.

The nearly static nature of the geometry during this ekpyrotic phase is underscored by the fact that only
a few e-folds of contraction are needed to generate 60 e-folds worth of perturbations. The comoving wavenumber for any given mode is related as usual to the moment of horizon crossing by $k\sim a|H|\sim (-t)^{p-1}$. Letting $i$ and $f$ denote respectively the initial and final time during which perturbations are generated, we therefore have
\be
e^{60} = \frac{k_f}{k_i}\sim\left(\frac{t_i}{t_f}\right)^{1-p}\sim \left(\frac{a_i}{a_f}\right)^{(1-p)/p}\,.
\ee
Since the spectral tilt is constrained observationally to be within 5\% of exact scale-invariance, we have $p\sim 1/40$, and therefore
\be
\ln\left(\frac{a_i}{a_f}\right)\approx 3/2\,.
\label{smallN}
\ee
Thus, in contrast with inflation where the universe grows exponentially big as the relevant range of modes is generated, here the universe only shrinks by a factor of order unity.

\subsection{Curvature Perturbation}

While $u$ is all we need to specify the scalar metric perturbations, it is useful to keep track of another gauge-invariant variable $\zeta$ --- the curvature perturbation on uniform-density hypersurfaces. This variable has the virtue of being constant on super-horizon scales, since there is no entropy perturbation in this single-field case. Thus, the large-scale spectrum of $\zeta$ calculated during the scaling phase automically agrees with the spectrum at horizon re-entry after reheating. 

Unfortunately, it turns out that the scale-invariant growing mode of $\Phi$ calculated above precisely projects out of $\zeta$. Since the two are related by
\be
\zeta = \frac{2}{3a^2(1+w)}\left(\frac{\Phi}{a'/a^3}\right)'\,,
\ee
it is easy to check using~(\ref{usoln}) and the background solution that the Newtonian potential precisely behaves as $\Phi\sim a'/a^3$ on large scales, leaving $\zeta$ with an unnacceptably strong blue tilt. This cancelation is at the core of all the controversy about the ekpyrotic spectrum of perturbations. 

While the derivation is rigorous, the conclusion is not without caveats. A crucial assumption is that the physics of the bounce from contraction to expansion all lies within the regime of validity of 4d effective theory, since the statement about $\zeta$ remaining constant relies on the equations of 4d gravity and matter. The possibility that higher-dimensional effects relevant at the bounce could imprint a scale-invariant contribution to $\zeta$ has been the subject of ample literature~\cite{tolley,other5d}.

In this work, however, we shall exploit an alternative possibility relying on two scalar fields and based on an observation by N.~Turok~\cite{talks} and work of Lehners, McFadden, Steinhardt and Turok~\cite{private}. This mechanism was suggested by~\cite{notari} and exploited in~\cite{finelli1,finelli2,finelli3}. Namely, with two scalar fields each rolling down their steep exponential potential, the entropy perturbation acquires a scale-invariant spectrum on large scales. If this can subsequently be imprinted onto $\zeta$, then $\zeta$ is endowed with a scale-invariant spectrum well before the bounce. After reviewing the mechanism of~\cite{private,talks}, we will show how the entropy perturbation imprints a scale-invariant contribution to $\zeta$ if one of the two fields reaches the minimum of its potential (region b) of Fig.~\ref{pot}) before the other. Since this field loses much kinetic energy in the process, this corresponds to a sharp turn in the trajectory in field space, and triggers a scale-invariant jump in $\zeta$. Then, by the same argument as above, whatever bounce physics comes into play to reverse contraction to expansion, as long as everything remains 4d and perturbative, the scale-invariant $\zeta$ will go through unscathed.

\section{Two-Field Ekpyrosis}
\label{2fields}

Remarkably, the scaling solution described above generalizes to two fields~\cite{private,finelli1}. Consider two scalars $\phi$ and $\psi$, each with a potential of the form shown in Fig.~\ref{pot}. In region (a) we can approximate the scalar potential as a sum of exponentials, generally with different powers $p\ll 1$ and $q\ll 1$:
\be
V(\phi,\psi) =  -V_0\exp\left(-\sqrt{\frac{2}{p}}\frac{\phi}{M_{\rm Pl}}\right) - U_0\exp\left(-\sqrt{\frac{2}{q}}\frac{\psi}{M_{\rm Pl}}\right)\,.
\label{V2field}
\ee
Furthermore, we continue to assume that each scalar field has canonical kinetic energy and that there are no higher derivative interactions. The dynamics of $\phi$ and $\psi$ are therefore governed by, respectively,
\bea
\nonumber
& & \ddot{\phi} + 3H\dot{\phi} = -V_{,\phi} \;;\\\
& & \ddot{\psi} + 3H\dot{\psi} = -V_{,\psi} \,.
\label{eoms}
\eea

When we come to perturbations, it will be useful to describe the field trajectory in $(\phi,\psi)$ space in terms of the adiabatic field $\sigma$, given by the geometrical relation
\be
\dot{\sigma} = \cos\theta \, \dot{\phi} + \sin\theta\, \dot{\psi} \,,
\label{dotsig}
\ee
where
\be
\tan \theta = \frac{\dot{\psi}}{\dot{\phi}}\,.
\ee
Combining~(\ref{eoms}), it is easily seen that $\sigma$ satisfies
\be
\ddot{\sigma} + 3H\dot{\sigma} = -V_{,\sigma}\,,
\label{sigmaeom}
\ee
where the slope of the potential along the field trajectory is just
\be
V_{,\sigma} = \cos\theta \, V_{,\phi} + \sin\theta \, V_{,\psi}\,.
\label{Vdsig}
\ee

In the exponential regime, there is an exact scaling solution to the Friedmann and scalar equations 
which generalizes~(\ref{scaling1}) to
\bea
\nonumber
& & a(t)\sim (-t)^{p+q}\,;\qquad H = \frac{p+q}{t}\,; \\
\nonumber
& & \phi(t) = 
\sqrt{2p}M_{\rm Pl}\log\left(-\sqrt{\frac{V_0}{M_{\rm Pl}^2p(1-3(p+q))}}\; t\right)\,; \\
& & \psi(t) = 
\sqrt{2q}M_{\rm Pl}\log\left(-\sqrt{\frac{U_0}{M_{\rm Pl}^2q(1-3(p+q))}}\; t\right)\,.
\label{scaling2}
\eea
Unlike the single-field case, however, the above solution is not an attractor because of an instability in the direction orthogonal to the field trajectory.
It is precisely this instability which is exploited in the next subsection to amplify entropy perturbations. Numerical analysis reveals that the instability grows fastest
along the direction of the steepest exponential. Suppose $p<q$ so that the potential for $\phi$ is steeper than that for $\psi$. Then, the instability brings $\psi$ to a halt, while the solution converges to the single-field scaling solution~(\ref{scaling1}) along the $\phi$ direction. Nevertheless, this is not of great concern since one only needs to be on the scaling solution for a few e-folds of contraction to generate the observable range of modes, as~(\ref{smallN}) demonstrates.

\subsection{Generation of Scale-Invariant $\delta s$ during the Scaling Regime} \label{dsgen}

In this subsection, we discuss the generation of a scale-invariant spectrum of entropy perturbations for the above scaling solution~\cite{private,finelli1}. See~\cite{gordon} for a nice review of entropy and adiabatic perturbations in multi-field models. A convenient gauge for the study of multi-field perturbations is the so-called spatially-flat or Mukhanov-Sasaki gauge~\cite{mukhsasaki}, in which the perturbations $\delta\phi$ and $\delta\psi$ in the two scalar fields satisfy the coupled equations
\bea
\nonumber
& & \ddot{\delta\phi} + 3H\dot{\delta\phi}+ \frac{k^2}{a^2} \delta\phi + \left\{V_{,\phi\phi}-\frac{1}{M_{\rm Pl}^2a^3}\frac{d}{dt}\left(\frac{a^3}{H}\dot{\phi}^2\right)\right\}\delta\phi
-\frac{1}{M_{\rm Pl}^2a^3}\frac{d}{dt}\left(\frac{a^3}{H}\dot{\phi}\dot{\psi}\right)\delta\psi = 0\,; \\
\nonumber
& & \ddot{\delta\psi} + 3H\dot{\delta\psi}+ \frac{k^2}{a^2} \delta\psi + \left\{V_{,\psi\psi}-\frac{1}{M_{\rm Pl}^2a^3}\frac{d}{dt}\left(\frac{a^3}{H}\dot{\psi}^2\right)\right\}\delta\psi
-\frac{1}{M_{\rm Pl}^2a^3}\frac{d}{dt}\left(\frac{a^3}{H}\dot{\phi}\dot{\psi}\right)\delta\phi = 0\,.
\label{coupled}
\eea
Thus, even though the scalar fields are uncoupled at the level of the potential, in the sense that $V_{,\psi\phi} = 0$, their perturbations become intertwined through gravity.

These fluctuations can be decomposed as a component along the field trajectory, $\delta\sigma$, which is the adiabatic perturbation, and a perturbation orthogonal to the trajectory, $\delta s$, which is the entropy or isocurvature perturbation:
\bea
\nonumber
\delta\sigma &=& \cos\theta\, \delta\phi + \sin\theta\,\delta\psi \\
\delta s &=& -\sin\theta\,\delta\phi + \cos\theta\, \delta\psi \,.
\label{deltafield}
\eea
Combining the ($\delta\psi$,$\delta\phi$) equations, and using the standard energy and momentum constraints, the entropy perturbation can be shown to satisfy~\cite{gordon}
\be
\ddot{\delta s} + 3H\dot{\delta s} + \left(\frac{k^2}{a^2}+V_{,ss}+3\dot{\theta}^2\right)\delta s = 4M_{\rm Pl}^2\frac{\dot{\theta}}{\dot{\sigma}}\frac{k^2}{a^2}\Psi\,,
\label{ds}
\ee
where $\Psi$ is the curvature perturbation in Newtonian gauge, while $V_{,ss}  = \cos^2\theta\, V_{,\psi\psi} + \sin^2\theta\, V_{,\phi\phi}$ is the curvature of the potential orthogonal to the field trajectory.

During the ekpyrotic scaling phase, described by~(\ref{scaling2}), both $\dot{\phi}$ and $\dot{\psi}$ scale as $1/t$ and, thus, $\theta$ is constant: 
\be
\tan\theta = \sqrt{\frac{q}{p}}\,.
\ee
Furthermore, for the scaling potential~(\ref{V2field}) we have
\be
V_{,ss} = -\frac{2}{t^2}\left(1-3(p+q)\right)\,.
\ee
Plugging all of this into~(\ref{ds}), we find
\be
\ddot{\delta s} + 3H\dot{\delta s} + \left\{\frac{k^2}{a^2} -\frac{2}{t^2}\left(1-3(p+q)\right)\right\}\delta s = 0\,.
\ee
This can be simplified further by introducing a rescaled variable $v = a\,\delta s$ and rewriting in terms of
conformal time, $d\tau = dt/a \sim dt\, t^{-(p+q)}$:  
\be
v'' + k^2v - \frac{2}{\tau^2}\left(1-\frac{3}{2}(p+q)\right) v \approx 0\,,
\label{v}
\ee
where we have dropped higher order terms in $p,q$. Here $\tau$ is assumed negative and increasing towards zero. Since $p$ and $q$ are small, the coefficient of the time-dependent term is approximately equal to 2, indicating a nearly scale-invariant spectrum. Indeed, the solution to~(\ref{v}) with standard Bunch-Davies initial conditions is, up to a phase,
\be
v = \frac{1}{\sqrt{2k}}\sqrt{\frac{\pi}{2}}\sqrt{-k\tau} H_{n}^{(1)}(-k\tau)\,,
\label{vsoln}
\ee
with
\be
 n \equiv \frac{3}{2}\sqrt{1-\frac{4}{3}(p+q)}\approx \frac{3}{2} - (p+q)\,.
\ee

On large scales, $k\ll aH$, the amplitude tends to $v \sim k^{-n}$, from which we can read off the spectral index:
\be
n_s - 1\approx 2(p+q) = 4\epsilon\,,
\label{nsexp}
\ee
where we have generalized the fast-roll parameter $\epsilon$ given in~(\ref{fastroll}) to the two-field case in the obvious way:
\be
\epsilon \equiv M_{\rm Pl}^{-2}\left(\frac{V}{V_{,\sigma}}\right)^2\ll 1\,.
\label{eps2field}
\ee
This parameter measures the steepness of the potential along the field trajectory.  Thus, for pure exponential potentials, the spectrum 
is slightly blue. More general potentials allow for red spectra as well, as we will show explicitly in the next section.

\section{Spectral Tilt for General Potentials} \label{red}

The small blue tilt for pure exponential potentials is a slightly disconcerting result in light of the recent 3-year WMAP data which favors a red tilt. However, we expect---and will show---
that this unambiguous blueness is an artifact of the pure exponential form. This intuition is supported by the single-field result: a non-zero
value for $\eta$ in~(\ref{single}) can change the sign of the tilt. We will see that the two-field story is nearly identical. 

More general potentials immediately imply departure from the scaling behavior studied earlier and thus generically require numerical analysis. For simplicity we
focus here on the case where both scalar fields have {\it identical} potential $\cV$ in region (a) of Fig.~\ref{pot}:
\be
V(\phi,\psi) = \cV(\phi) + \cV(\psi)\,.
\ee
This approach was originally mentioned in~\cite{talks}, and some of the results below
overlap with~\cite{private}. Of course the potentials should only be identical in region (a), not globally, for
otherwise the entropy perturbation will never get converted to the adiabatic mode. We can assume, for instance, that the minimum in region (b) occurs at different field values. 

If the fields start out with identical initial conditions, then their time-evolution will also be the same, and thus $\theta = \pi/4$. It follows that the adiabatic field introduced in~(\ref{dotsig}) satisfies $\dot{\sigma} = ( \dot{\phi}+ \dot{\psi})/\sqrt{2}$. Similarly for the slope of the potential along the field trajectory, given by~(\ref{Vdsig}): $V_{,\sigma} = \left( \cV_{,\phi} + \cV_{,\psi}\right)/\sqrt{2}$.
Thus the Friedmann and ``$\dot{H}$" equations can be expressed as
\bea
\nonumber
& & 3H^2M_{\rm Pl}^2 = \frac{1}{2}\dot{\sigma}^2 + V\,; \\
& & \dot{H}M_{\rm Pl}^2 = -\frac{1}{2}\dot{\sigma}^2\,.
\label{backsigma}
\eea
Combined with the $\sigma$ equation of motion~(\ref{sigmaeom}), this rewriting makes manifest the virtue of having identical potentials for the two scalars --- all background equations of motion reduce effectively to that of a single effective scalar field $\sigma$. 

Moving on to perturbations, the evolution equation~(\ref{ds}) for the entropy mode greatly simplifies because $\dot{\theta}=0$. Moreover, we can rewrite the time-dependent potential term, $V_{,ss}$, in terms of $\sigma$-derivatives: $V_{,ss}  = \cos^2\theta\, \cV_{,\psi\psi} + \sin^2\theta\, \cV_{,\phi\phi} = V_{,\sigma\sigma}$. Our starting point for the generation of $\delta s$ is therefore
\be
\ddot{\delta s} + 3H\dot{\delta s} + \left(\frac{k^2}{a^2}+V_{,\sigma\sigma}\right)\delta s = 0\,.
\label{dsnew}
\ee

Since the background no longer follows some exact scaling solution, the way to proceed is to recast every term in the equation
solely in terms of the background equation of state parameter,
\be
\bar{\epsilon} \equiv \frac{3}{2}(1+w) = -\frac{\dot{H}}{H^2} = -\frac{d\ln H}{dN}\,,
\ee
and its derivatives. Here $N\equiv \ln a$ is the number of e-folds, as usual. (In the limit of pure exponential potentials, $\bar{\epsilon}$ is of course constant and related to the $\epsilon$ parameter in~(\ref{eps2field}) by $\bar{\epsilon} = 1/2\epsilon$.) 

To illustrate the method, let us combine equations~(\ref{backsigma}) to obtain a relation for the potential:
\be
V = H^2M_{\rm Pl}^2\left(3-\bar{\epsilon}\right)\,.
\label{Veps}
\ee
Then, from the second of~(\ref{backsigma}) and the definition of $\bar{\epsilon}$, we immediately have $d\sigma/dN = \sqrt{2\bar{\epsilon}}M_{\rm Pl}$, which allows 
rewriting the derivative of the potential as
\be
V_{,\sigma}  = -H^2M_{\rm Pl}\sqrt{2\bar{\epsilon}}\left(3-\bar{\epsilon} +\frac{1}{2}\frac{d\ln\bar{\epsilon}}{dN}\right)\,.
\label{dVeps}
\ee
Similarly, we have
\be
\frac{V_{,\sigma\sigma}}{H^2} = -2\bar{\epsilon}^2 + 6\bar{\epsilon}+\frac{5}{2}\frac{d\bar{\epsilon}}{dN}-\frac{3}{2}\frac{d\ln \bar{\epsilon}}{dN}-\frac{1}{4}\left(\frac{d\ln \bar{\epsilon}}{dN}\right)^2 - \frac{1}{2}\frac{d^2\ln \bar{\epsilon}}{dN^2}\,.
\ee

Inspired by analogous calculations in the inflationary context~\cite{wang}, it is convenient to rewrite~(\ref{dsnew}) in terms of a dimensionless time variable
\be
x \equiv \frac{1}{\bar{\epsilon}-1}\;\frac{k}{aH}\,,
\ee
and, as we did in Sec.~\ref{dsgen}, rescale the perturbation variable to
\be
v \equiv  a\; \delta s\,.
\ee
After some algebra, the perturbation equation~(\ref{dsnew}) takes the exact final form
\bea
\nonumber
& & \left(1-\frac{1}{\bar{\epsilon}-1}\frac{d\ln(\bar{\epsilon}-1)}{dN}\right)^2x^2\frac{d^2v}{dx^2} + \frac{1}{(\bar{\epsilon}-1)^2}\left[\left(\frac{d\ln(\bar{\epsilon}-1)}{dN}\right)^2-\frac{d^2\ln(\bar{\epsilon}-1)}{dN^2}\right]x\frac{dv}{dx} + x^2v \\
& &  \;\;\;+  \frac{1}{(\bar{\epsilon}-1)^2}\left\{-2  -2\bar{\epsilon}^2 + 7\bar{\epsilon}+\frac{5}{2}\frac{d\bar{\epsilon}}{dN}-\frac{3}{2}\frac{d\ln \bar{\epsilon}}{dN}-\frac{1}{4}\left(\frac{d\ln \bar{\epsilon}}{dN}\right)^2 - \frac{1}{2}\frac{d^2\ln \bar{\epsilon}}{dN^2}\right\}v = 0\,.
\label{vfinal}
\eea

As advocated, this depends only the background equation of state parameter and its time derivatives. To simplify things, we use the fact $\bar{\epsilon}$ is large during the ekpyrotic phase, corresponding to $w\gg 1$. Moreover, we assume it is a slowly-varying function of $N$ over the observable range of modes. To leading order, it follows from~(\ref{Veps}) and~(\ref{dVeps}) that
\be
\bar{\epsilon}\approx \frac{1}{2\epsilon}\,,
\ee
where $\epsilon$ was defined in~(\ref{eps2field}). Similarly, it is easy to show that
\be
\frac{d\ln\bar{\epsilon}}{dN} = 4\bar{\epsilon}\eta\,,
\label{epsder}
\ee
where we have generalized the $\eta$ parameter of~(\ref{fastroll}) to the two-field case:
\be
\eta \equiv 1 - \frac{V_{,\,\sigma\sigma}V}{V_{,\,\sigma}^2}\,.
\ee

In the approximation that $\epsilon$ and $\eta$ are small and nearly constant,~(\ref{vfinal}) collapses to the simple form 
\be
x^2\frac{d^2v}{dx^2}  + \frac{x^2}{1-8\eta}v - 2\left(1 - 3(\epsilon-\eta)\right)v = 0\,,
\ee
where the coefficient of $x^2$ can be reabsorbed by a constant rescaling of the time variable. This final form can be recast as a Bessel equation, as before, and the corresponding
spectral tilt can be read off immediately
\be
n_s-1\approx 4\left(\epsilon - \eta\right)\,,
\label{ns}
\ee
As a check, this agrees with~(\ref{nsexp}) in the limit of pure exponential potentials. 

More importantly, we see that a red tilt is possible if $\eta$ is positive and dominates over the $\epsilon$ term. This mild condition is satisfied by a wide class of potentials. For instance, any potential of the form $V(\phi) \sim \exp(-\phi^n)$, with $n>2$, gives a red tilt at large $\phi$.

\section{Converting Entropy to the Adiabatic Mode} \label{imprint}

Next we turn to the conversion of the scale-invariant entropy mode to the curvature perturbation. Indeed, the upshot is that the entropy perturbation sources $\zeta$ on large scales:
\be
\dot{\zeta} \approx -2H\frac{\dot{\theta}}{\dot{\sigma}}\delta s\,.
\label{dotzeta}
\ee
If the field trajectory is a straight line, so that $\theta$ remains constant, then $\zeta$ remains constant as well. If, however, the field trajectory has curvature or makes a sharp turn, then $\dot{\zeta}$ will change on large scales. In the context of our scenario, we present an explicit mechanism for converting the nearly scale-invariant entropy perturbation into the adiabatic mode, which naturally exploits the desired shape of the potential shown in Fig.~\ref{pot}. 

During the scaling solution, region (a) of the potential, $\delta s$ acquires a scale-invariant spectrum on large scales, as described earlier. Since this scaling solution corresponds to $\theta={\rm const.}$, however, it follows from~(\ref{dotzeta}) that $\zeta$ is {\it not} scale-invariant, as in the single-field ekpyrotic model. Eventually the potentials for $\phi$ and $\psi$ must eventually depart from pure exponential by hitting a minimum and growing to positive values, corresponding to region b). The scaling solution abruptly ends.  In general, one of the two fields, say $\psi$, will hit the minimum before the other. As $\psi$ climbs up the steep hill, its kinetic energy decreases tremendously, while that of $\phi$ stays nearly the same. Hence, in field space this corresponds to a sharp turn in the trajectory, from $\theta\approx \arctan\sqrt{q/p}$ to $\theta\approx 0$. This jump in $\theta$ implies a jump in $\zeta$ jumps as well, which thereby acquires a scale-invariant spectrum from the entropy perturbation.

If the rise in the $\psi$ potential is sufficiently steep, then the change in $\theta$ occurs almost instantaneously compared to a Hubble time. This rapid-transition approximation is consistent 
with the fact that, in the scaling regime, we have $\dot{\psi}\gg H$ since $q\ll 1$, as seen from~(\ref{scaling2}). Thus, if the rise in the potential is indeed sufficiently steep, the field climbs on the plateau of region c) in a very short time compared to the Hubble time. Thus we can rewrite~(\ref{dotzeta}) as
\be
\dot{\zeta} \approx -\frac{2H}{\dot{\sigma}}\arctan\left(\sqrt{\frac{q}{p}}\right)\delta(t-t_i)\,\delta s \,,
\label{dotzeta2}
\ee
where we have denoted the time of the transition by $t_i$. We can argue that each factor in this expression, save for the delta-function term, is approximately constant during the transition.
This is obviously the case for $H$, by assumption. To see that the same is true of $\dot{\sigma}$, let us look back at~(\ref{sigmaeom}). As the field trajectory makes a sharp turn, the slope of the potential $V_{,\sigma}$ will generically be discontinuous, {\it i.e.}, have a theta-function jump. If $\dot{\sigma}$ were to jump as well, however, the $\ddot{\sigma}$ term in~(\ref{sigmaeom}) would generate a delta-function in the equation of motion, which is inconsistent. Hence $\dot{\sigma}$ must be continuous, and we can therefore substitute its value of $\sqrt{2(p+q)}/t_i$ just before the transition. Finally, $\delta s$ does change, but by at most a factor of order unity. The proof requires some steps, which we will provide at the end of the section. For the moment let us treat it as essentially constant.

Making use of this rapid-transition approximation, we can integrate~(\ref{dotzeta2}) and find that the curvature perturbation $\zeta$ inherits a nearly scale-invariant spectrum from $\delta s$:
\be
\left|\zeta\right| \approx \frac{2H}{\dot{\sigma}}\arctan\left(\sqrt{\frac{q}{p}}\right) \delta s\,,
\ee
with spectral tilt given by~(\ref{nsexp}). Ignoring the change of order unity of during the transition, we can substitute the large-scale $\delta s$ by taking 
the limit $k\rightarrow 0$ of~(\ref{vsoln}), neglecting for simplicity the small departure from scale-invariance:
\be
k^{3/2}\delta s \approx \frac{1}{\sqrt{2}a(-\tau)} \approx \frac{-H}{\sqrt{2}(p+q)}\,.
\label{dssoln}
\ee
Substituting this, as well as the scaling solution $\dot{\sigma} = \sqrt{2}H M_{{\rm Pl}}/\sqrt{p+q}$, 
we obtain
\be
k^3\zeta^2 \approx \frac{H^2}{2\epsilon M^2_{{\rm Pl}}} \arctan^2\left(\sqrt{\frac{q}{p}}\right)\,,
\ee
where $\epsilon$ is the fast-roll parameter defined in~(\ref{eps2field}). Remarkably, up to the trigonometric factor of order unity, this is nearly identical to the corresponding expression in slow-roll inflation: 
\be
k^3\zeta^2_{\rm inf} \sim \frac{H^2}{\epsilon_{\rm inf} M^2_{\rm Pl}}\,, 
\ee
where $\epsilon_{\rm inf} \equiv M_{\rm Pl}^2 (V_{,\phi}/V)^2/2$ is the standard slow-roll parameter. 
This confirms earlier expectations~\cite{design} that the level of 
tuning on the ekpyrotic potential from the amplitude of density 
perturbations is comparable to its inflationary counterpart.

It remains to prove that the entropy perturbation $\delta s$ does not change dramatically during this process. Since the modes of interest are already far outside the horizon, we can study~(\ref{ds}) in the limit $k\rightarrow 0$:
\be
\ddot{\delta s} + 3H\dot{\delta s} + \left(V_{,ss}+3\dot{\theta}^2\right)\delta s = 0\,.
\label{ds2}
\ee
For concreteness we consider the case where $\psi$ hits the minimum of its potential, climbs up a steep hill and hits a flat plateau, as described above. Since $\psi$ loses most of its kinetic energy in the process, we have $\theta\rightarrow 0$. Moreover, $V_{,\psi\psi}\rightarrow 0$ since the potential becomes flat. Hence, $V_{,ss}\rightarrow 0$ as well. Finally, since the transition is assumed nearly instantaneous on a Hubble time, we can safely ignore the Hubble damping term. Thus~(\ref{ds2}) reduces to:
\be
\ddot{\delta s} \approx -3\dot{\theta}^2\delta s\,.
\label{ds3}
\ee

Here the approximation $\dot{\theta}\sim \delta(t)$ made earlier is too drastic since one would have to make sense of $\delta^2(t)$. Thus we have to be a more precise about
$\theta(t)$ during the transition in order to solve for $\delta s$. To proceed analytically, let us model the transition by a linear extrapolation between the initial and final angles, $\theta_i$ and $\theta_f$, respectively:
\be
\theta(t) = \theta_i + \frac{(t-t_i)}{T}(\theta_f-\theta_i)\,.
\ee
Thus the transition begins at $t=t_i$ and ends at $t=t_i+T$, with $HT\ll 1$. In the end we will choose $\theta_i\approx \arctan\sqrt{q/p}$ and $\theta_f\approx 0$, corresponding to the case of interest, but for now let us be general. The solution to~(\ref{ds3}) is then given by
\be
\delta s (t) = (\delta s)_i\cos\left(\frac{\sqrt{3}(\theta_f-\theta_i)(t-t_i)}{T}\right) + \frac{T}{\sqrt{3}(\theta_f-\theta_i)}(\dot{\delta s})_i\sin\left(\frac{\sqrt{3}(\theta_f-\theta_i)(t-t_i)}{T}\right)\,,
\ee
where $(\delta s)_i$ and $(\dot{\delta s})_i$ are respectively the amplitude and time-derivative of $\delta s$ at the onset of the transition. From (\ref{dssoln}) we see that $(\dot{\delta s})_i\sim H(\delta s)_i$, and therefore $(\dot{\delta s})_i T\ll (\delta s)_i$. Hence the final amplitude of the entropy perturbation is given by
\be
(\delta s)_f \approx (\delta s)_i\cos\left(\sqrt{3}(\theta_f-\theta_i)\right)\,.
\ee
For the case of interest, $|\theta_f-\theta_i| \approx \arctan\sqrt{q/p}$, which proves our claim that $\delta s$ changes by a factor of order unity during the transition.

\section{General Remarks on Bouncing Cosmologies} \label{remarks}
 
Next we turn to the all-important issue of reversing from contraction to expansion in a non-singular fashion. Now that $\zeta$ has been shown to be scale-invariant in the contracting phase, whatever physics generates a bounce, the usual conservation of $\zeta$ on super-horizon scales guarantees that the universe will emerge in the expanding phase endowed with a
scale-invariant spectrum. 

Unfortunately there is certainly not  a wealth of consistent bouncing scenarios to choose from. The recently proposed mechanism of Creminelli {\it et al.}~\cite{Ghost3} based on ghost condensation~\cite{Ghost1,Ghost2} is at present the {\it only} such mechanism all within the realm of a consistent effective field theory and in which all instabilities can be kept under control. There are some concerns as to whether something like ghost condensation can be realized in a consistent theory of quantum gravity, such as string theory~\cite{IRnima}. But for the purpose of this scenario we will not be worried about issues of UV completion. The essential point is that it provides a consistent and ghost-free effective theory.

Let us begin by making a few general remarks about obtaining a successful bounce. Since $H< 0$ during a contracting phase, by definition, in order to have a bounce ($H=0$)
we need some epoch during which $\dot{H}>0$. But this is highly non-trivial to achieve in any particle physics model. Consider for instance a non-linear sigma model of $N$ scalar fields with arbitrary potential~\cite{seiberg}:
\be
{\cal L} = \frac{1}{2}G_{ij}\left(\phi^k\right)\partial\phi^i\partial\phi^j + V\left(\phi^i\right)\,.
\ee
As long as the metric on moduli space $G_{ij}$ is positive-definite, then a bounce is simply impossible, since
\be
\dot{H} = -\frac{M_{\rm Pl}^2}{2}G_{ij}\left(\phi^k\right)\partial\phi^i\partial\phi^j\leq 0\,,
\ee
independently of the potential. More generally, for a perfect fluid with energy density $\rho$ and 
pressure ${\cal P}$, we have
\be
M_{\rm Pl}^2\dot{H} = -\frac{1}{2}\left(\rho+{\cal P}\right)\,.
\label{dotH}
\ee
Thus a necessary condition for a bounce is a violation of the NEC: $\rho+{\cal P}\geq 0$. 
(In covariant form, the NEC is the requirement that $T_{\mu\nu}n^\mu n^\nu \geq 0$ for any 
null vector~\cite{HE}.)

Of course one can get $\dot{H}>0$ by relaxing the positive-definiteness of $G_{ij}$, {\it i.e.} allowing for ghost-like modes. But ghosts have disastrous consequences for the viability of the theory. 
In order to regulate the rate of vacuum decay one must invoke explicit Lorentz breaking at some low scale~\cite{Cline}. In any case there is no sense in which a theory with ghosts can be thought as an effective theory, since the ghost instability is present all the way to the UV cut-off of the theory. More generally, it was argued in~\cite{nic} that in generic 2-derivative theories, violations of the NEC immediately imply the presence of ghosts or tachyons with arbitrarily-large mass. The ghost condensate evades this theorem since it relies on higher-derivative kinetic terms, while nevertheless defining a consistent effective theory. 

Even if some component violates the NEC without introducing catastrophic instabilities, it is still a non-trivial feat to dynamically achieve a bounce. The point is that in order for $\dot{H}$ and then $H$ to reverse sign, this NEC-violating component must come to dominate the energy. Since the equation of state is $w<-1$, by definition, this means that its energy density scales like
\be
\rho_{\rm NEC} \sim a^\alpha\,,
\ee
where $\alpha$ is some {\it positive} power. In an expanding universe, this grows while everything else redshifts, meaning that this component is bound to dominate the universe. But in a contracting universe, of course, precisely the opposite happens. While everything else blueshifts, $\rho_{\rm NEC}$ becomes negligible. Put another way, any bouncing solution is unstable to the addition of energy into a normal component --- in order to achieve a bounce, one must ensure that nearly all the energy somehow gets funneled to the NEC-violating fluid. This is a key challenge for any scenario of bouncing cosmology. 

In the remaining sections of the paper, we show explicitly how the ghost-condensate bounce can be successfully merged with the preceding ekpyrotic phase. The outcome is a singularity-free bouncing cosmology in which a scale-invariant spectrum of perturbations is generated during the contracting phase and transferred unscathed through the bounce. 


\section{Essentials of Ghost Condensation}
\label{ghostcond}


Theories of ghost condensation describe a scalar field with higher-derivative kinetic term~\cite{Ghost1,Ghost2}
\begin{equation}
{\cal{L}}= \sqrt{-g} M^{4}P(X),
\label{evgeny} 
\end{equation}
where 
\be
X  = -\frac{1}{2m^4}(\partial\phi)^2
\ee
is dimensionless, and $M$ and $m$ are some arbitrary scales 
to be determined by the fundamental theory. (The ghost condensate literature 
usually defines $X=-(\partial\phi)^2/2$, with $\phi$ having dimension of length. 
Here, however, we stick to the usual mass dimension for scalars, for consistency with earlier sections.) Theories of the form~(\ref{evgeny}) were first introduced in cosmology as $k$-inflation~\cite{kinflation} or $k$-essence models~\cite{kessence} to drive accelerated expansion without potential energy.

In a cosmological context, the scalar satisfies
\be
\frac{d}{dt}\left(a^3P_{,\,X}\dot{\phi}\right) = 0\,.
\label{Xeom}
\ee
For generic $P(X)$, this implies the usual redshifting (blueshifting) 
of scalar kinetic energy as the universe expands (contracts). 
However, if $P(X)$ displays a minimum at some finite $X$, which by rescaling of $m$ can be 
chosen to lie at $X_0 = 1/2$, then $X=1/2$ provides an exact solution 
to~(\ref{Xeom}). This corresponds to the field maintaining {\it constant} 
kinetic energy and thus growing linearly in time
\be
\phi(t) = -m^2t\,.
\label{linearphi}
\ee
(Of course this is a solution for any extremum of $P(X)$. However, as we will see shortly, fluctuations around a maximum are ghost-like.) 
The purely derivative nature of the Lagrangian~(\ref{evgeny}) is technically natural if there is some global shift symmetry $\phi \rightarrow \phi + const$.
Moreover, since the above solution is linear in time, corrections involving more than one time derivatives on $\phi$, such as $\left(\Box\phi\right)^2$, vanish identically. Thus,
near a minimum of $P(X)$,~(\ref{evgeny}) provides a consistent effective field theory. 

The full stress tensor of the ghost condensate is derived as usual from~(\ref{evgeny})
\be
T_{\mu\nu} = g_{\mu\nu}M^4P(X) + P_{,\,X}\partial_\mu\phi\partial_\nu\phi\,,
\ee
corresponding in the cosmological context to an energy density and pressure given by
\bea
\nonumber
\rho &=& M^4\left(2P_{,\,X}X-P\right)   \\
{\cal P} &=& M^4P(X)\,.
\eea
In particular, at an extremum of $P$ the ghost condensate behaves as a fluid 
with $w=-1$, mimicking the effect of a cosmological term.

Ignoring gravity for a moment, small fluctuations in the scalar field, 
$\phi = -m^2t+ \pi$, around a constant $X$ solution have the quadratic lagrangian
\be
{\cal L}\propto \left(P_{,\,X} \left(X_0\right)+2X_0P_{,\,XX}\left(X_0\right)\right)\dot{\pi}^2 - P_{,\,X} \left(X_0\right)(\vec{\nabla}\pi)^2\,. 
\label{fluc1}
\ee
In particular, around an extremum of $P(X)$, perturbations have ``right-sign" kinetic term if the extremum is a {\it minimum}, and are ghostlike for a maximum. Furthermore, the
gradient term vanishes at that point, meaning that one must keep the leading correction to the $P(X)$ lagrangian, say from $\left(\Box\phi\right)^2$, to obtain
\be
{\cal L}_{\rm grad}\sim -\frac{1}{M'^2}(\vec{\nabla}^2\pi)^2\,.
\ee
This implies the dispersion relation $\omega^2  \sim  k^4/M'^2$. (To make contact with the normalization convention in~\cite{Ghost3}, their $\bar{M}$ is related to our $M'$ by $M'=M^2/\bar{M}$.) We postpone the stability analysis of small fluctuations, including metric perturbations, to Sec.~\ref{instab}. We first describe how NEC violation is achieved in ghost condensation.


\subsection{Violating the NEC}


At the minimum of $P(X)$, the ghost 
condensate behaves as a cosmological term and as such is just at the borderline 
of violating the NEC. To push it in the $w<-1$ region, 
following~\cite{Ghost3} we introduce a potential $V(\phi)$ for the scalar. 
If this potential is sufficiently flat,  
then it gives a small correction to the scalar equation of motion, 
allowing us to treat it as a small perturbation to~(\ref{linearphi}): 
$\phi = -m^2t+\pi$. To leading order in $\pi$, the energy density and pressure become
\bea
\nonumber
\rho &\approx & -\frac{K M^2 \dot{\pi}}{m^2}+V \\
{\cal P} &\approx & -V\,,
\eea
where $K\equiv P_{,\,XX}(1/2)$ is the (dimensionless) curvature of the 
kinetic function about the ghost condensate point. 
Meanwhile we have set $P(1/2) = 0$ through a constant shift in $V$.

And here we discover the culprit for NEC violations 
in ghost condensate, in the form of a term {\it linear} 
in $\dot{\pi}$ in the energy density. Indeed, the $\dot{H}$ equation --- see~(\ref{dotH}) 
 --- takes the form
\be
M_{\rm Pl}^2\dot{H} =  \frac{1}{2}\frac{K M^4\dot{\pi}}{m^2}\,,
\label{dotHpi}
\ee
and therefore can have either sign, depending on the dynamics of $\pi$. 
The latter is determined by 
expanding $\frac{M^4}{m^4}a^{-3}\partial_t\left(a^3P_{,\,X}\dot{\phi}\right) = -V_{,\,\phi}$ 
around the $\phi = -m^2t$ solution:
\be
\ddot{\pi} + 3H\dot{\pi} = -\frac{V_{,\phi}}{K}\frac{m^4}{M^4}\,.
\label{pieom}
\ee
This looks like the equation of motion for a scalar field, except that 
here $V_{,\,\phi}$ is to zeroth order independent 
of $\pi$ --- instead it is a time-dependent source term 
determined by the background $\phi = -m^2t$ evolution.


\subsection{Taming the Instabilities} \label{instab}


The ghost condensate model suffers from two types of instabilities: one of the 
Jeans-type which is present even in standard ghost condensation; and a second, 
gradient-type instability which is due to the NEC violating background. 
We briefly review these in turn, starting from the latter. 
First consider expanding the original $P(X)$ action~(\ref{evgeny}) around the NEC-violating background
\be
\phi = -m^2t + \pi_0(t) + \pi(t,\vec{x})\,,
\ee
where $\pi_0(t)$ satisfies~(\ref{pieom}). 
Using that $\dot{H} \sim \dot{\pi}$ from~(\ref{dotHpi}), the result is~\cite{Ghost3}
\be
{\cal L} \propto \frac{1}{2}\dot{\pi}^2 + 
\frac{\dot{H}M_{\rm Pl}^2}{KM^4} (\vec{\nabla}\pi)^2 - 
\frac{m^4}{2 K M^4 M'^2}(\vec{\nabla}^2\pi)^2\,.
\label{fluc2}
\ee
Thus a non-vanishing $\dot{H}$ generates a non-zero gradient 
term for the fluctuations, which 
evidently has the wrong sign on the NEC-violating background.  Correspondingly,  the
dispersion relation,
\be
\omega^2 = - 2\frac{\dot{H}M_{\rm Pl}^2}{KM^4}k^2 + \frac{m^4}{K M^4 M'^2}k^4\,,
\ee
has a gradient instability at long wavelengths. The rate of this instability 
peaks at $k^2\sim \dot{H}M_{\rm Pl}^2M'^2/m^4$, corresponding to
\be
\omega_{\rm grad}^2 \sim \left(\frac{\dot{H}M_{\rm Pl}^2M'}{\sqrt{K}M^2 m^2}\right)^2\,.
\ee
Of course such instabilities are absent if their rate is less than the 
Hubble rate, $|\omega_{\rm grad}| \; \lsim \; |H|$, giving the bound
\be
\frac{\dot{H}}{H}\;\lsim\; \frac{\sqrt{K} M^2 m^2}{M'M_{\rm Pl}^2}\,.
\label{bound1}
\ee

The second type of instability arises when including mixing of $\pi$ with gravity and is present even in standard ghost condensation. One finds a Jeans-type instability with
rate~\cite{Ghost1,Ghost2}
\be
\omega_{\rm Jeans} \sim \frac{\sqrt{K}M^2 m^2}{M'M_{\rm Pl}^2}\,,
\ee
which once again is harmless if less than the Hubble rate:
\be
\frac{\sqrt{K}M^2 m^2}{M'M_{\rm Pl}^2} \;\lsim \; H\,.
\label{bound2}
\ee

Equations~(\ref{bound1}) and~(\ref{bound2}) 
together require that $M$ and $M'$ be chosen such that they fit in the range
\be
\frac{\dot{H}}{H}\;\lsim\; \frac{\sqrt{K} M^2 m^2}{M'M_{\rm Pl}^2} \;\lsim \; H\,.
\label{comb}
\ee
Evidently in order to have a parametric window for which this can be satisfied, 
it must be that $\dot{H}\ll H^2$. If $\dot{H}\sim H^2$, for instance, 
then~(\ref{comb}) implies a relation
between $M$, $m$ and $M'$; whereas $\dot{H}\gg H^2$ implies 
that one of the instabilities is present. As we approach the 
bounce and $|H|\rightarrow 0$, clearly the condition $\dot{H}\ll H^2$ must break 
down sooner or later. Since the instabilities of interest are of the Jeans-type, 
however, their effects can be mitigated if the entire period of 
NEC-violation lasts roughly one e-fold, so that
\be
H\Delta t \;\lsim \; 1\,,
\label{duration}
\ee
where $H$ is the Hubble parameter at the onset of NEC-violation, say. 


\section{Merging Ekpyrosis and Ghost Condensate} \label{merge}


The next step consists of merging the ekpyrotic phase with the 
NEC-violating ghost condensate to reverse contraction to expansion in a
non-singular fashion. For simplicity we focus first on the case of 
single-field ekpyrosis --- we will discuss the generalization to two fields
at the end of the section.

As mentioned in Sec.~\ref{remarks}, the existence of an NEC-violating fluid 
does not guarantee a bounce --- it must also come to dominate the
universe in order to reverse the sign of $\dot{H}$. But this is a 
non-trivial feat since its energy density redshifts instead of 
blueshifting as the universe contracts.
Thus, for instance, if the ghost condensate is a spectator field 
different than the scalar relevant for the ekpyrotic  
phase, one must ensure that somehow the energy in the former
gets efficiently transfered to the latter, with negligible energy going to 
other light degrees of freedom. This seems to us a highly unnatural possibility.

Instead if the ekpyrotic and ghost condensate scalars are one of the same 
field --- let us denote it again by $\phi$ ---, then the transfer of energy is of
course a non-issue. We will describe what this merging entails for the 
global form of $P(X)$ and $V(\phi)$, and derive
consistency conditions required for a successful bounce. As we will see, 
these relations can be satisfied for a wide class of kinetic functions and scalar potentials. 

Each subsection deals with different aspects of this merger. For pedagogical 
purposes, we provide at the conclusion of each subsection a short summary of the key results.


\subsection{Global Form of $P(X)$}


Unlike ghost inflation, where the inflaton is in the ghost condensate 
phase while driving inflation, here it is crucial that the scalar field $\phi$ has approximately 
standard kinetic term in the ekpyrotic phase, {\it i.e.} in region a) of Fig.~\ref{pot}. That is, 
\be
M^4P\left(X\right) \approx m^4 X =
-\frac{1}{2}(\partial\phi)^2 \qquad ({\rm ekpyrotic}\;{\rm phase})\,.
\label{Pek}
\ee
This follows immediately from the fact that $V$ is steep in this region and 
therefore would generically violate the flatness condition of ghost condensate.
Furthermore the perturbation calculation crucially relies on the kinetic term being canonical.

Of course in the ghost phase, $P(X)$ must have some minimum 
where the field can condensate. In its vicinity, we can assume the following quadratic
form:
\be
P(X) \approx \frac{K}{2}\left(X-\frac{1}{2} \right)^2 \qquad ({\rm ghost}\;{\rm phase})\,,
\label{Pghost}
\ee
so that $\phi =-m^2 t$ at the minimum while $P_{,\,XX} = K$, consistent 
with the conventions in our earlier discussion.

\begin{figure}[ht]
\centering
\includegraphics[width=150mm]{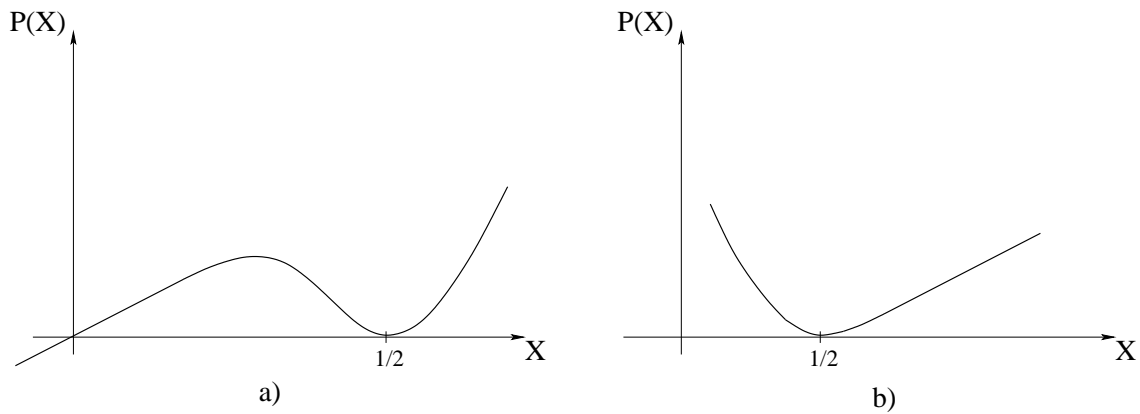}
\caption{Two possible choices for the global behavior of the kinetic function $P(X)$. 
In case a), the linear regime (ekpyrotic phase) lies at smaller $X$ than the 
ghost condensate point. This necessarily
implies a maximum for $P(X)$ in between, signaling the presence of a real ghost.  
In case b), the linear regime lies at larger $X$, thereby avoiding real ghosts.}
\label{Pchoice}
\end{figure}

One might expect that the quasi-linear behavior~(\ref{Pek}) occurs for 
small $X$, while the minimum in~(\ref{Pghost}) is generated when 
higher-derivative corrections become
important at sufficiently large $X$. This picture would also be consistent 
with the fact that $X$ increases in the ekpyrotic phase, due to cosmological blueshift. 
However, a moment's thought
reveals that such a $P(X)$ would necessarily have a {\it maximum} somewhere 
in between, thereby signaling the existence of a real ghost. 
This is sketched in Fig.~\ref{Pchoice}a. If one is willing to 
tolerate a real ghost, then of course there is no need to invoke all of the ghost condensate technology.

So let us instead assume that the quasi-linear regime occurs at larger $X$ than the ghost condensate point, as sketched in Fig.~\ref{Pchoice}b. Thus, at the onset of the ekpyrotic phase, the scalar kinetic energy starts out to the right of the minimum, say, and moves away from it. Thus the challenge here is to bring the field back to the vicinity of the minimum of $P(X)$ at the end of the ekpyrotic phase. As we will see this can be achieved naturally by the form of the potential. We will not attempt to motivate this form of $P(X)$ from a UV-complete theory. Instead our goal is to provide a specific example of $P(X)$ which, combined with the general form of the potential shown in Fig.~\ref{pot}, yields a non-singular alternative cosmology to inflation.

To summarize, we have argued that having a canonical form for the kinetic term during the ekpyrotic phase, combined with the no-ghost constraint during the subsequent evolution towards the ghost condensate point, requires the kinetic function $P(X)$ to have the form shown in Fig.~\ref{Pchoice}b.


\subsection{Scalar Potential $V(\phi)$}

\begin{figure}[ht]
\centering
\includegraphics[width=150mm]{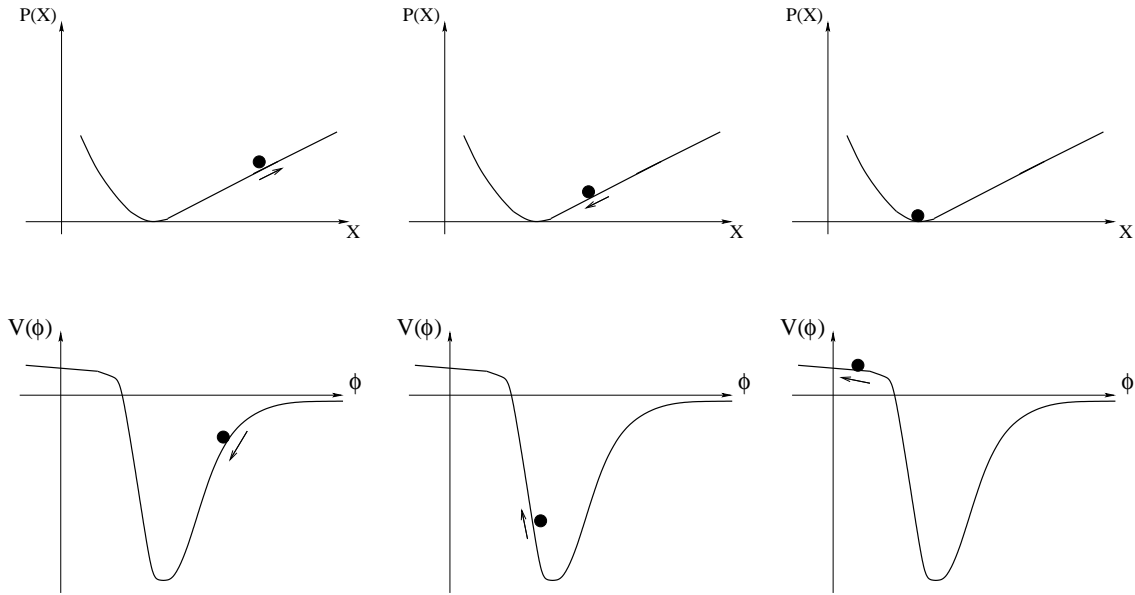}
\caption{Sketch of the dynamical evolution of the scalar field, in terms of $V$ and $P(X)$.}
\label{cartoon}
\end{figure}

Since $X$ grows during the contracting phase, it must somehow find its way back to the vicinity of the minimum of $P(X)$ at the end of the ekpyrotic phase. The idea is to exploit the
sharp rise in the potential, shown in region~(b) of Fig.~\ref{pot}, to greatly reduce the kinetic energy in the scalar and bring $X$ near the ghost condensate point. 
In other words, by appropriately choosing the difference in potential energy between the value at the minimum and that on the plateau, the field will lose just the right amount of kinetic energy to bring $X$ near the minimum of $P(X)$. The field evolution in both potential and kinetic function space is sketched in parallel in Fig.~\ref{cartoon}.

We can be more precise about the form of the potential in region c) during the ghost condensate phase. Since the latter requires a flat potential,
we can approximate $V(\phi)$ as linear in $\phi$:
\be
V(\phi) \approx \alpha\Lambda^4
\left(1-\beta\frac{\Lambda^2}{m^2}\frac{\phi}{M_{\rm Pl}}\right)\,,
\label{Vlin}
\ee
where $\Lambda$ is some scale, while $\alpha$ and $\beta$ are dimensionless. 
The form of the $\beta$-term is chosen for convenience, as will become apparent shortly. 

As we now argue, $\alpha$ and $\beta$ must both be {\it positive}. 
To prove the former one only needs to look at the Friedmann equation, which, 
to leading order in $\pi$, is just
\be
3H^2M_{\rm Pl}^2 = -\frac{K M^4\dot{\pi}}{m^2} +V 
\approx -\frac{K M^4\dot{\pi}}{m^2} +\alpha\Lambda^4\,.
\label{fried}
\ee
But since $\dot{\pi} > 0$ during the NEC-violating phase, as seen in~(\ref{dotHpi}), 
it follows that $\alpha>0$. We can therefore set  $\alpha=1$ without loss of generality. 

To argue that $\beta$ must be positive as well is equally straightforward. 
What we want is for the NEC to be initially satisfied ($\dot{\pi} < 0$) 
and then violated for some time ($\dot{\pi}>0$). Precisely at the onset of NEC violation, 
however, corresponding $\dot{\pi}=0$,~(\ref{pieom}) reduces 
to $\ddot{\pi} = \beta \Lambda^6 m^2/KM^4M_{\rm Pl}$. Hence to proceed to the NEC-violating phase, 
we need $\beta>0$. Note that a potential of the form~(\ref{Vlin}) 
with $\alpha,\beta>0$ is indeed consistent with region c) of Fig.~\ref{pot}. 

In this subsection we have shown that the scalar potential in the 
ghost condensate regime must be positive, flat, and have negative slope. 
The last condition follows from our choice that $\dot{\phi} < 0$ at the 
ghost condensate point, to be consistent with the field motion during the ekpyrotic phase.


\subsection{Dynamics of the Bounce}


A key approximation that must hold throughout the NEC-violating phase and 
through the bounce for perturbation theory to apply is
\be
\dot{\pi}\ll m^2\,.
\label{condpi}
\ee
To see what this entails on the potential and kinetic function, 
let us combine~(\ref{dotHpi}),~(\ref{Vlin}) and~(\ref{fried}) to obtain the master equation 
\be
M_{\rm Pl}^2\left(3H^2 + 2\dot{H}\right)  = \Lambda^4\left(1 + \frac{\beta\Lambda^2}{M_{\rm Pl}} t\right) \approx \Lambda^4\,,
\label{master}
\ee
valid to leading order in $\dot{\pi}$.  In the last step we have further assumed that  $\beta\Lambda^2 \Delta t/M_{\rm Pl} \;\lsim \; 1$ for the period of interest, as we will prove shortly. Now, since $\dot{\pi} \sim \dot{H}$, the condition~(\ref{condpi}) is most stringent when $\dot{H}$ is maximal. Evidently this occurs at the bounce itself, when $H=0$, and thus $\dot{H}\approx \Lambda^4/2M_{\rm Pl}^2$. Translated in terms of $\dot{\pi}$, it follows that a necessary condition for~(\ref{condpi}) to be satisfied is
\be
\Lambda^4 \ll M^4K\,.
\label{lambcond}
\ee
But then, looking back at~(\ref{fried}), this immediately implies that the 
Hubble rate is also constrained throughout the ghost condensation phase:
\be
H \ll \frac{\sqrt{K}M^2}{M_{\rm Pl}}\,.
\ee

There is one further condition, namely~(\ref{duration}), having to do with the suppression of Jeans-like instabilities during the bounce. 
To see what it implies, let us introduce the dimensionless parameters
\be
\tilde{H}=H \frac{M_{\rm Pl}}{\Lambda^2}\,, \qquad \tilde{t}=t\frac{\Lambda^2}{M_{\rm Pl}}\,,
\label{master2}
\ee
in terms of which~(\ref{master}) takes the simplified form
\be
2\dot{\tilde{H}}+3\tilde{H}^2=1+\beta\tilde{t}\,.
\label{master3}
\ee

\begin{figure}
\epsfxsize=4.5in \epsffile{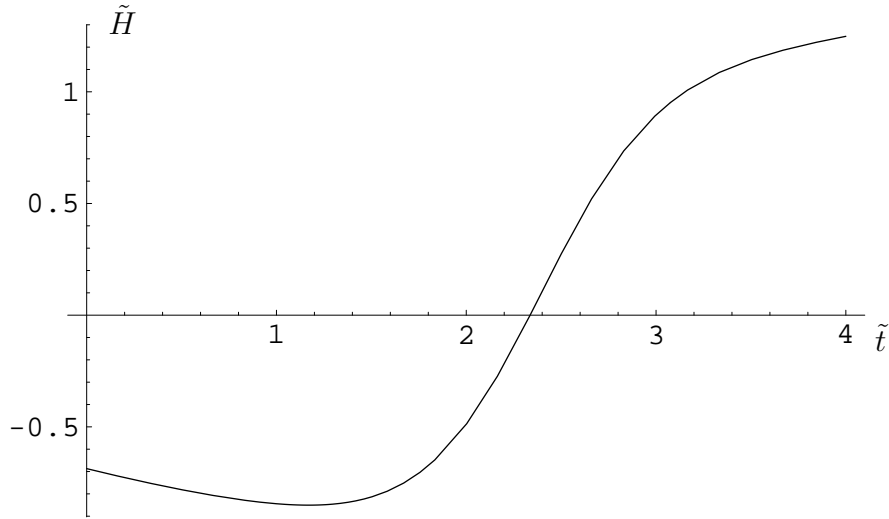}
\begin{picture}(30,30)
\put(-290,190){$\tilde{H}$}
\put(0, 70){$\tilde{t}$}
\end{picture}
\caption{Evolution of $\tilde{H}(\tilde{t})$ for $\beta=1$. We see that first $\dot{\tilde{H}}$ 
vanishes at some time, 
and later on $\tilde{H}$ itself vanishes, marking the reversal from contraction to expansion.
Note that, in solving~\eqref{master3}, initial conditions were chosen such that 
$t=0$ corresponds to the beginning of the ghost condensation phase.}
\label{Fig4}
\end{figure}

In terms of these dimensionless variables, the condition~(\ref{duration}) 
is just $\tilde{H}\Delta\tilde{t}\;\lsim\; 1$. Thus, provided that
\be
\beta\sim \cO(1)\,,
\ee
then there is only one time scale in~(\ref{master3}), and thus~(\ref{duration}) is guaranteed to hold. Figure~\ref{Fig4} shows the result of numerically integrating~(\ref{master3}) for $\beta=1$.
The initial conditions are such that $\dot{H}\ll H^2$ at the beginning, ensuring that all instabilities are under control. These conditions lead to a NEC-violating phase, followed by a bounce. Since $\beta\sim\cO(1)$, the duration of the NEC-violating phase is indeed of order one in dimensionless time units, satisfying the requirement~(\ref{duration}).
%

Knowing that $\beta\sim\cO(1)$, we can determine the 
change in $\phi$ during the NEC-violating  phase:
\be
\Delta\phi = m^2\Delta t\sim \frac{m^2}{H}\sim \frac{m^2}{\Lambda^2}M_{\rm Pl}\,,
\label{zh3}
\ee
where in the third step we have estimated the Hubble parameter during the 
ghost condensate phase by its approximate value at the onset of NEC violation: 
$H\sim \Lambda^2/M_{\rm Pl}$ --- see~(\ref{master}). 
We will see further on that $m$ has to be small, therefore~(\ref{zh3}) 
generically implies that $\phi$ moves by a small distance compared to the Planck mass
during the ghost condensate phase.
Incidentally, note from~(\ref{Vlin}) that the variation 
of $V$ is then of order $\Delta V \sim \Lambda^4$, and thus $V$ changes by a 
factor of order one in the process. This justifies the approximation made in~(\ref{master}).

To summarize, perturbation theory around the ghost condensate is valid 
throughout provided that the scale $M$ of the kinetic function is much greater
than that of the potential as well as the Hubble parameter. 
Furthermore, in order to keep gradient instabilities under control, 
we have shown that the parameter $\beta$
characterizing the slope of the potential must be of order unity. 
These conditions are easily satisfied for a wide class of potentials and kinetic functions. 
It was explicitly shown that this yields a non-singular bounce.


\subsection{Consistency Relations for a Successful Merger}


Let us now focus on the transition as the field passes through region~(b). 
Since this corresponds to the transition between the ekpyrotic phase and the ghost-condensate bouncing
phase, we will find consistency relations between the kinetic 
function $P(X)$ and the scalar potential $V(\phi)$.

As we did when discussing perturbations, we model this transition as instantaneous compared to a Hubble time. This is justified since, just before reaching $V_{\rm min}$, the field obeys the scaling solution and, therefore, is moving rapidly on a Hubble time, as shown by~(\ref{scaling1}):
\be
\frac{1}{2M_{\rm Pl}^2}\dot{\phi}^2 = \frac{H^2}{p}\gg H^2\,.
\ee
If the rise in $V$ is sufficiently sharp, then $\phi$ will indeed reach the plateau in a very short time compared to the Hubble time. In this case we can treat $H$ as constant during this transition, by energy conservation. Let us denote it by $H_{\rm min}$. Then, from~(\ref{scaling1}) we obtain
\be
M^2_{\rm{Pl}}H_{\rm min}^2 = -\frac{p}{1-3p}V_{\rm min}\approx -pV_{\rm min}\,.
\label{Hmin}
\ee
Of course this is expected since $\dot{\phi}^2/2\approx -V$ during the ekpyrotic phase. 

Once we reach the plateau, the assumption is that the field is near 
the ghost condensate point. We will see that this 
constrains $V_{\rm min}$ to lie within some parametric window. 
First, to derive an upper bound, we start from the Friedmann 
equation~(\ref{fried}) at the onset of the ghost phase:
\be
3H^2_{\rm min}M_{\rm Pl}^2 \approx -\frac{K M^4\dot{\pi}}{m^2} + \Lambda^4\,.
\ee
Substituting~(\ref{Hmin}) and rearranging, we get
\be
\dot{\pi}\approx m^2\left(\frac{pV_{\rm min}}{KM^4}+\frac{\Lambda^4}{KM^4}\right)\,.
\ee
Now a consistency condition for being near the ghost condensate point is 
that $\dot{\pi}\ll M^2$, as we recall from~(\ref{condpi}). 
Since we already know that $\Lambda^4 \ll KM^4$ from~(\ref{lambcond}), it immediately follows that
\be 
\left\vert V_{\rm min}\right\vert \ll \frac{M^4K}{p}\,.
\label{upper}
\ee

To derive a lower bound, note that by assumption we must have 
$\dot{\phi}^2 \; \gg \; m^4$ during the ekpyrotic process,
since $P(X)$ has to be approximately linear corresponding to canonical 
kinetic energy. 
However, we also 
know that $\dot{\phi}^2\approx -2V$ during the scaling phase, and therefore
\be
\left\vert V \right\vert \; \gg \; m^4\,
\label{lower1}
\ee
throughout ekpyrosis. 
Since at this phase $V$ is monotonous we have to require
\be
\left\vert V_{\rm{ek}} \right\vert \; \gg \; m^4\,,
\label{lower2}
\ee
where $V_{\rm{ek}}$ is the value of the potential at the onset of the ekpyrosis. Let us relate 
$V_{\rm{min}}$ and $V_{\rm{ek}}$ through the number of e-folds ${\cal N}$. Since during  
ekpyrosis the scale factor $a$ stays approximately constant, the number of e-folds is 
\begin{equation}
e^{{\cal N}}=\frac{H_{\rm{min}}}{H_{\rm{ek}}}. 
\label{zh1}
\end{equation}
Hence, from the solution~(\ref{scaling1}) it follows that 
\begin{equation}
\left\vert V_{\rm ek} \right\vert =e^{-2{\cal N}}\left\vert V_{\rm min} \right\vert. 
\label{zh2}
\end{equation}

Equations~(\ref{upper}), (\ref{lower2}) and~(\ref{zh2}) 
together imply that $V_{\rm min}$ must lie within the range
\be
m^4e^{2{\cal N}} \;\ll\; \left\vert V_{\rm min}\right\vert  \;\ll\; \frac{M^4K}{p}\,.
\label{range}
\ee
Noting that ${\cal N}$ is 60 or so, this requires $m$ to be less than $M$ by many orders of magnitude.
Since these two scales are physically different, the above allowed range can be substantially broad.
Indeed, from the above considerations, $M$ denotes the cutoff scale by which all higher-derivative terms are suppressed,
whereas $m$ sets the expectation value of $\dot{\phi}$ at the minimum of $P(X)$.
Incidentally, the upper bound is, in fact, conservative since it follows from 
the relatively strict requirement that the field be near the ghost condensate point by the
time it reaches the plateau. More general transitions will therefore loosen this bound.

To summarize, in order for the field to land in the vicinity of the ghost condensate 
point at the end of the ekpyrotic phase, the minimum of the potential, $V_{\rm min}$, must
lie within the above range. 
The lower bound comes from the requirement that the kinetic energy of $\phi$ is approximately canonical 
and, thus, is large comparing to $m$. 
On the other hand it cannot be too large.  
Otherwise by the time $\phi$ reaches the plateau its kinetic function $P(X)$ will not be 
near the condensation point. 
Since $m$, $M$, $V_{\rm min}$ and $K$ are free parameters at the level of model-building, 
we see that these conditions are satisfied for a wide class of models.


\subsection{Generalization to Two Fields}


We conclude with a few words on the generalization to two fields, which is straightforward.
Essentially both $\phi$ and $\psi$ are assumed to have qualitatively the same $P(X)$ and the same qualitative shape for their
potentials. Both fields will reach their respective $V_{\rm min}$ more or less at the same time and will proceed to their respective
ghost condensate point. (Of course these transitions cannot happen simultaneously since our mechanism for converting
entropy to adiabatic perturbations relies on one field reaching $V_{\rm min}$ before the other. Nevertheless this time delay could
be negligible compared to a Hubble time, for instance.) Once both fields have reaches their ghost condensate points, both act as NEC-violating fluids and drive the universe
towards a non-singular big bounce.

\section{Discussion of Reheating}
\label{reheat}

After a non-singular bounce has been successfully completed, the energy in the ghost condensate must somehow get converted into matter and radiation degrees of freedom in order to reheat the universe. We briefly comment on two possible reheating mechanisms, one entirely at the level of a 4d effective theory, the other inspired by
the original ekpyrotic scenario and concepts of heterotic M-theory.

Let us start with the 4d effective mechanism, which has been mentioned in the ghost inflation context~\cite{ghostinflation}. After the bounce, most of the energy of the ghost condensate is stored in its potential energy --- the kinetic energy is proportional to $2P_{\,,X}X-P$, which is small near the ghost condensate point. To trigger reheating, one assumes that at some field value the scalar potential displays a precipitous drop towards zero, after which it becomes flat again. When the field reaches the drop, its kinetic energy will be greatly perturbed away from the ghost condensate point. If $\phi$ is coupled to matter fields, this non-adiabatic process will excite matter and radiation degrees of freedom. In other words, the potential energy difference reheats the universe. At the end of the reheating phase,
the scalar field settles back to the ghost condensate point, but with negligible residual energy.
Reheating by this mechanism is easily achieved within the context of the new ekpyrotic scenario.

Much more speculative, the second mechanism is inspired by the original ekpyrotic scenario.  Here,
the ekpyrotic scalar field has the geometrical interpretation of the distance between a bulk M5 brane 
and the observable 
end-of-the-world boundary brane. Reheating occurs when the bulk brane 
inelastically collides and is absorbed by the boundary brane. This fusion proceeds through
a ``small instanton" transition, during which the nature of the light degrees of 
freedom can change; greatly increasing, for example, the number of massless scalar fields at the collision.
In the context of new ekpyrosis, higher-derivative 
corrections to the M5-brane modulus kinetic term could generate a ghost 
condensate point, allowing for a non-singular bounce before the collision. 
The subsequent brane collision then excites these scalars which, in turn, 
transfer their energy to matter and radiation and reheat the universe. 
In this context, it is natural to expect that the ghost condensate becomes 
massive and disappears following the small instanton transition.

\section{Conclusion}
\label{conclude}

In this paper, we have presented a new and fully consistent scenario for the origin of the primordial density perturbations. Instead of being generated through a rapid phase of accelerated
expansion shortly after the big bang, here the perturbations are generated in a phase of slow contraction, long before the big bang. The key breakthrough in this paper is a non-singular bouncing
cosmology, achieved by successfully merging the ekpyrotic phase with a subsequent NEC-violating ghost condensate phase.  We have derived the explicit consistency relations required for this merger
to be successful. These can be fulfilled by a wide class of kinetic functions and scalar potentials. The entire cosmological evolution is, therefore, under control and can be
tracked throughout at the level of a 4d effective theory. 

This framework allows us to settle the controversial issue of the fate of perturbations through the bounce. A non-singular bounce allows perturbations to remain in
the linear regime throughout. More importantly, since the evolution can be described within a 4d effective theory, the curvature perturbation, $\zeta$,
is unambiguously conserved and goes through the bounce unscathed. To generate a scale-invariant spectrum for $\zeta$ in the pre big bang phase, we
have made use of a recently proposed mechanism of entropy perturbation generation~\cite{private,talks}. This is accomplished by having two ekpyrotic scalar fields rolling down their respective
negative exponential potentials. 
We have found that the resulting amplitude for $\zeta$ is remarkably similar to its counterpart in inflation. This confirms earlier claims in the ekpyrotic literature that the required level of tuning
on the ekpryotic and inflationary scalar potentials from COBE normalization is comparable. Additionally, the spectral index is slightly red for a wide class of potentials, consistent with recent evidence from WMAP.

As with inflation, the remaining challenge is to embed this scenario 
within a UV complete theory of quantum gravity, such as string theory. 
Potentials of the type required in new ekpyrosis can be realized in $N=1$ supergravity,
including low energy effective string theories.
However, the question of whether one can realize ghost condensation in string theory
remains open. Ultimately the bounce could be generated using an
entirely different mechanism. And indeed many of the results described here, 
such as the generation of a scale-invariant $\zeta$ and its propagation through the 
bounce, would apply equally well. Our motivation in focusing on the ghost 
condensate was to provide as concrete a realization of our scenario as possible.

At the level of a cosmological scenario,  
``New Ekpyrotic Cosmology" provides a consistent alternative paradigm to inflationary cosmology.
The two scenarios make distinctive predictions for the gravitational wave spectrum: the 
inflationary spectrum is nearly scale-invariant, whereas that of ekpyrotic cosmology is very blue
and, therefore, unobservable on large scales~\cite{gwaves}.
Moreover, the generic prediction of the simplest inflationary models is a significant gravity wave amplitude, just below the current sensitivity levels of microwave
background experiments~\cite{compare}.  Ekpyrosis,  on the other hand, 
predicts an unobservably small amplitude. Thus the failure to detect B-mode polarization in upcoming experiments would place inflation in an uncomfortable corner~\cite{compare}, while lending support to the ekpyrotic paradigm. \\

{\bf Acknowledgments}
We are most grateful to S.~Dubovsky, P.~Creminelli, A.~Nicolis, P.J.~Steinhardt and N.~Turok for helpful discussions. 
This work of B.A.O. is supported in part by the DOE under contract No. DE-AC02-76-ER-03071 and by the NSF Focused Research Grant DMS0139799. 
The research of E.I.B. and J.K. at Perimeter Institute is supported in part by the Government of Canada through NSERC and by the Province of Ontario through MRI. \\



\end{document}